# Polarized superradiance from delocalized exciton transitions in tetracene single crystals

A. Camposeo[1]*, M. Polo[1,2], S. Tavazzi[3], L. Silvestri[3], P. Spearman[3], R. Cingolani[1] and D. Pisignano[1,2]

[1] *National Nanotechnology Laboratory of CNR-INFM, c/o Distretto Tecnologico, Università del Salento, via Arnesano I-73100, Lecce, Italy*

[2] *Scuola Superiore ISUFI, Università del Salento, via Arnesano I-73100, Lecce, Italy*

[3] *Dipartimento Scienza dei Materiali, Università di Milano Bicocca, Via Cozzi 53, I-20125 Milano Italy*

## Abstract

Polarized superradiant emission and exciton delocalization in tetracene single crystals are reported. Polarization-, time-, and temperature-resolved spectroscopy evidence the complete polarization of the zero-phonon line of the intrinsic tetracene emission from both the lower ($F$ state) and the upper (thermally activated) Davydov excitons. The superradiance of the $F$ emission is substantiated by a nearly linear decrease of the radiative lifetime with temperature, being fifteen times shorter at 30 K compared to the isolated molecule, with an exciton delocalization of about 40 molecules.

* Corresponding author. Tel: +390832298147. Fax: +390832298146. *E-mail address*: andrea.camposeo@cnr.it

Superradiance is a cooperative process resulting in coherent light from a domain of correlated emitters, behaving like a single quantum-mechanical system. The oscillator strength of the ensemble is enhanced with respect to the single emitter and scales with the size ($N_s$) of the superradiant domain leading to a shortening of the radiative lifetime and a line narrowing of the transition. A few classes of polycrystalline organic materials exhibit superradiance [1-4]. For instance, Meinardi et al. [1] inferred superradiance in quaterthiophene (4T) thin films by a net increase of the radiative decay rate of the zero-phonon (0-0) line at temperatures ($T$) around 5 K.

In organic *single crystals*, the occurrence of superradiance is forecasted by models originally developed for two-dimensional herringbone systems [5-7]. The emission is attributed to a thermalized population of dressed excitons originating from the coupling between excitons from the lowest molecular optical transition and a phonon band from a single strongly coupled intramolecular vibrational mode. For aggregates with non-vanishing component of the molecular transition dipole moment in the herringbone plane, a (0-0) superradiant transition is predicted. This should be polarized as the lower Davydov exciton, and its intensity should scale as $N_T/N_{th}$, where $N_T$ indicates the temperature dependent number of coherently emitting molecules in the aggregate and $N_{th}$ is the threshold for superradiance. $N_{th}$ is determined by the lattice arrangement of the molecules and by the projection of the molecular transition dipole moment on the herringbone plane. However, despite the importance and wide use of organic single crystals for applications [8-10], the experimental evidence of superradiance in such systems, which should manifest a well-defined polarization state and larger coherence domain, is not yet unequivocally resolved.

Tetracene (TCN), an organic semiconductor widely studied for its electronic properties [11], is a particularly interesting system to probe superradiance. It consists of four fused benzene rings and crystallizes in a layered herringbone structure, whose unit cell is triclinic. The cell includes two non-equivalent molecules under translational operation. The molecular electronic transition at the lowest energy is polarized along the short molecular axis (*M*) [12], which has a strong component in the *ab* crystal plane (accessible face for optical measurements). Differently from *J*- and *H*-type aggregates, the herringbone arrangement of the molecules in the *ab* plane gives rise to two allowed excitonic bands [2,13], the lowest in energy being the most intense, polarized along different directions and with non-zero components along all the crystallographic axes, although with major components in the *ab* plane. Light polarized along the axis (*x*) of maximum absorption in the *ab* plane mainly comes from the lower Davydov state [13]. In particular, the molecular arrangement enables the occurrence of superradiance without a threshold size [2]. Superradiance was first observed in TCN nanoaggregates by Lim et al., finding a four-fold decrease of the radiative lifetime at 4 K compared to isolated molecules [2], and recently on thin films grown on highly oriented pyrolitic graphite by Voigt et al. [3]. Since the photogeneration and emission processes change considerably depending on the amount of structural disorder, as evidenced for the singlet-triplet exciton relaxation in TCN and pentacene [14], the study of single crystals is necessary to assess the origin and the polarization of all the emission transitions, and the observation of coherent effects on larger domains.

In this Letter, we report a detailed analysis of the emission of TCN single crystals and of its polarization states, for *T* ranging between 11 K and 290 K. Polarized and time-resolved spectroscopy allow us to distinguish the intrinsic excitonic emission, providing the first evidence of the strong polarization of both the superradiant (0-0) excitonic transition and the thermally activated (0-0) excitonic transition from the upper Davydov component, in contrast to the other emission bands. Upon decreasing *T*, the radiative decay rate increases up to fifteen times the isolated molecule rate. The data show remarkable agreement with a model developed by Spano [2,7], concerning both the superradiant behaviour of the (0-0) transition and its polarization.

TCN is purchased from Sigma-Aldrich and purified by sublimation cycles. The purified powder is placed in a glass crucible and heated at 170 °C in a crystal growth system by the physical vapour transport method, using nitrogen as gas carrier. Single crystals, selected by inspection under a polarizing optical microscope, are placed on quartz substrates (1×1 cm$^2$) for optical characterization. In order to rule out re-absorption of the emitted photons, which occurs mainly at the band edge due to the overlap of the absorption and emission spectra, and polarization mixing effects [13], a crystal with thickness 70 nm is selected for optical investigation [15]. The cw photoluminescence (PL) is excited by a polarized diode laser ($\lambda$=405 nm and power <100 µW), normally incident on the *ab* face with a 1 mm excitation spot, and collected by a fiber-coupled monochromator equipped with a charge-coupled device (CCD). The room-temperature PL absolute quantum yield ($\Phi$) is measured by placing the sample in an integrating sphere [16]. PL measurements at variable *T* are performed by mounting the samples in a He closed-cycle cryostat under vacuum ($10^{-4}$ mbar), and collecting the forward emission from the irradiated surface, i.e. along the normal to the crystal *ab* face. Time-resolved PL measurements are performed by exciting the crystal with the second harmonic of a mode-locked Ti:Sapphire laser, delivering pulses of duration <100 fs at about 800 nm. The PL signal is dispersed by a monochromator coupled to a streak camera equipped with a two-dimensional CCD, providing wavelength- and time-resolved data, with time resolution of about 20 ps. No dynamics or spectral changes in the time integrated spectra are observed by varying the average incident power in the range 0.1-10 mW. The measurements are performed with a average laser power of 120 µW and a diameter of the focussed laser of about 0.1 mm.

The polarized cw PL spectra of a TCN single crystal at 290 K, collected by aligning the polarization analyzer parallel and perpendicular to the *x* axis, are displayed in Fig. 1a. The emission at room temperature is characterized by an intense *x*-polarized peak at 530 nm, corresponding to the (0-0) emission of the lowest excitonic transition, followed by replicas at about 565 nm (0-1) and 615 nm (0-2). In the high energy tail of the spectrum with polarization perpendicular to *x* a small shoulder is observed at 510 nm, which corresponds to the (0-0) emission line of the upper Davydov exciton (Fig. 1b). The (0-0) transition peaked at 530 nm has a polarization contrast $\Psi_x = (I_{//x} - I_{\perp x})/(I_{//x} + I_{\perp x}) = 0.95$ ($I_{//x}$ and $I_{\perp x}$ are the intensity of the PL with polarization analyzer parallel and perpendicular to the *x* axis, respectively), whereas the peak at 510 nm is polarized perpendicularly to *x* ($\Psi_x < -0.4$) and the 0-1 and 0-2 replicas are only partially polarized.

These findings are well interpreted on the basis of models [7] which predict an intense polarized (0-0) line originating from the lowest energy excitonic state and a weak, thermally activated, polarized (0-0) line originating from the upper Davydov exciton (having energy higher than the lower Davydov exciton by $E_{DS}$ = 0.08 eV) [13], together with partially polarized replicas. At 290 K, the ratio of the (0-0) emission intensities of the lower and upper Davidov states is predicted to be $I_{0-0}^{Low}/I_{0-0}^{Up} = 3.8 \times \exp(E_{DS}/k_B T) \approx 90$, where the 3.8 coefficient represents the ratio of the oscillator strengths of the two transitions calculated from measured absorption spectra [13]. This value is in satisfactory agreement with the experimental value of 50 deduced from Fig. 1b, confirming our assignment of the 510 nm peak to the (0-0) upper Davydov emission. The presence of the (0-0) transition of the upper Davydov state is not reported in previous studies of the PL of TCN crystals [17,18], because of the presence of strong spectrally overlapping self-trapped states. The emission at low temperature (Fig. 1c) is characterized by the presence of a sharp (linewidth 150 cm$^{-1}$) peak at 526 nm, corresponding to the (0-0) excitonic transition, followed by peaks at 539 and 558 nm. As for wavelengths above 550 nm the emission is due to the convolution of trap states and vibronic replicas [3,17], we focus our analysis on the first two peaks. We can clearly identify the (0-0) transition at 11 K, differently from previous studies [17], where the (0-0) excitonic state at low $T$ was detectable as an unresolved shoulder in the PL high energy region, probably due to self absorption and/or crystal quality.

In Fig. 1d, we display the emission intensity vs the analyzer angle with respect to the $x$ axis, for the peaks at 526 and 539 nm. The latter exhibits a lower polarization contrast ($\Psi_{x,539}$ = 0.90 vs $\Psi_{x,526}$ = 0.98) and its polarization axis is rotated by 4° with respect to the (0-0) excitonic transition. TCN crystals are reported to undergo a structural phase transition upon cooling [19]. The triclinic polymorph stable at high temperature (HT) changes into a low temperature (LT), also triclinic, polymorph, having slightly different lattice parameters and a smaller volume of the unit cell [19-21]. A red-shift (300-600 cm$^{-1}$) of the HT exciton transition is reported, suggesting that the two phases possess different (0-0) emitting free excitons, generally referred as $F$ and $F'$, respectively [22]. The origin of the *F'* emission in TCN thin films is found to be due to a dynamical relaxation process occurring during the first 50 ps [3]. Our analysis of the time-resolved emission in TCN single crystals evidences the existence of a long-life ($\tau_{PL} \sim$ 3 ns) and broad (linewidth at 11 K of 340 cm$^{-1}$) state at 539 nm, (Fig. 2), whose PL lifetime and intensity increase upon decreasing *T* below 90 K. Based on these observations and the possibility of having self-trapping of the exciton in TCN single crystals [17], we attribute *F'* to a defect state, as also supported by the lack of hysteresis in the emission spectra [15] upon a cooling/heating cycle. A similar behaviour was recently observed in anthracene [23], where the singlet exciton is coupled to a trapped state. Thus, we make use of a three level model [23] to analyze our polarized time-resolved PL data, aiming to determine $\tau_{PL}$ for the intrinsic (0-0) transition [15]. The radiative lifetime ($\tau_r$) at room temperature (12 ns), obtained by the measured PL quantum yield, $\Phi = \tau_{PL}/\tau_r = 0.013$, and the transition emission lifetime $\tau_{PL}$ ($\tau_{PL}^{-1} = \tau_r^{-1} + \tau_{nr}^{-1}$, where $\tau_{nr}$ is the non radiative lifetime) turns out to be shorter than that of the single molecule (29-32 ns) [2]. The dependence of $\tau_r$ on *T* can be derived by rescaling the room temperature value by the exciton emission intensity ($I_{0-0}$, proportional to $\Phi$) and its lifetime. Upon lowering the temperature $\tau_r$ decreases almost linearly down to a value of 1.8 ns (Fig. 3), much smaller than the single molecule value, indicating the superradiant behaviour of the (0-0) transition. The linear dependence is typical of two-dimensional systems [2]. Accordingly, the single crystal radiative decay rate ($k_{crystal}$) shows a fifteen-fold enhancement at 30 K with respect to the corresponding monomer rate ($k_{monomer}$) and a temperature dependence similar to the theoretical predictions for a planar aggregate [2] (inset of Fig. 3). Since $k_{crystal} / k_{monomer}$ is proportional to $N_T$, based on the comparison with the calculated values for a 4×4 aggregate (for which $N_{T=0}$=16) we estimate an exciton delocalization of about 40 molecules at 30 K.

Further evidence of (0-0) superradiance comes from the temperature behaviour of the (0-0) transition in the temperature range where *F'* is not seen. The inset of Fig. 4 shows the emission spectra integrated in the first 200 ps. The (0-0) emission exhibits a remarkable increase upon decreasing $T$ as compared to the (0-1) vibronic replica which remains almost constant, in agreement with theoretical predictions [7]. From the spectral areas we measure a $I_{0-0}/I_{0-1}$ ratio of 2 at room temperature, increasing to 5 at 110 K, larger than the value observed for the monomer in solution (~1) [2].

Our results demonstrate the possibility of obtaining larger coherence volumes in organic single crystals compared to molecular polycrystalline systems [2], limitations being mainly due to disorder and crystal quality. In order to investigate this aspect, we analyze the (0-0) transition linewidth (full width at half maximum, *FWHM*), which narrows considerably from 600 cm$^{-1}$ at room temperature to 150 cm$^{-1}$ at 50 K, and levels off as the temperature lowers further (Fig. 4). We attribute this behaviour to the presence of static disorder which prevents further delocalization of the excitation, in agreement with the results of Fig. 3 where the radiative decay rate is shown to start levelling off at about 50 K.

In summary, the results allow to conclude that the emission in TCN single crystals mainly originates from the lowest Davydov state. In particular, we have experimental evidence of the polarization of the zero-phonon line of the intrinsic TCN emission (*F* state) and of a large increase of its radiative decay rate upon decreasing $T$, a signature of its superradiant nature. The single crystal radiative lifetime is about two times shorter at room temperature and fifteen times shorter at 30 K with respect to the isolated molecule, indicating an unprecedented level of exciton delocalization, estimated to be about 40 molecules. The demonstration of polarized superradiance in TCN single crystals opens the way to the use of superradiant emission from molecular materials with microcavity mode selection [24], possibly leading to even larger coherent domains.

**Acknowledgments**

This work was supported by the Italian Ministry of University and Research PRIN program. We acknowledge Prof. A. Papagni for TCN purification, Dr. M. Campione for growing crystals, Dr.

**Figure captions**

**Figure 1.** Polarized PL spectra of a TCN single crystal, collected by the analyzer axis parallel (continuous blue line) or perpendicular (continuous green line) to the $x$ axis, at 290 K (a) and 11 K (c). (b) High energy part of the polarized PL spectra at 300 K. The arrows indicate the (0-0) transition from the lower (LOW) and upper (UP) Davydov state. (d) Emission intensity at 11 K of the peaks at 526 nm (blue dotted line) and 539 nm (red continuous line) as a function of the analyzer angle. 0° corresponds to the analyzer axis parallel to the $x$ axis.

**Figure 2.** $x$-polarized, time-resolved PL spectra at 11 K. Time-resolved PL spectra are integrated over time intervals of a few tens of ps and vertically shifted for better clarity. Inset: Time-evolution of the emission intensity from the states at 526 nm (red continuous line) and 539 nm (blue dashed line) at 11 K.

**Figure 3.** Radiative lifetime, $\tau_r$, of the (0-0) line vs. $T$. The dashed line is a fit to the data in the temperature interval 11-290 K. Inset: Radiative decay rate of the (0-0) line normalized by the monomer radiative decay rate (empty circles), compared with calculated values for a 4×4 TCN aggregate (dotted curve, from Ref. [2]).

**Figure 4**. Emission linewidth vs $T$ for the (0-0) line measured from cw $x$-polarized PL spectra (full squares) and time-resolved, $x$-polarized, spectra integrated in the first 50 ps (empty circles). Inset: PL spectra integrated in the first 200 ps of emission, measured at different $T$. From bottom to top, $T$ decreases from 290 to 110 K.

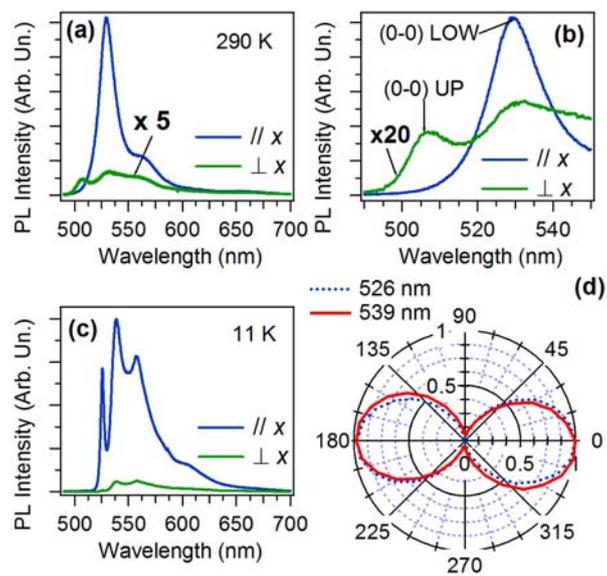

Figure 1, A.Camposeo et al.

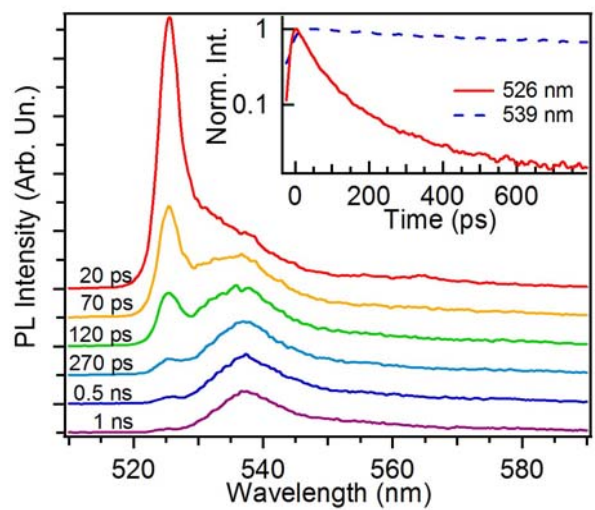

Figure 2, A.Camposeo et al.

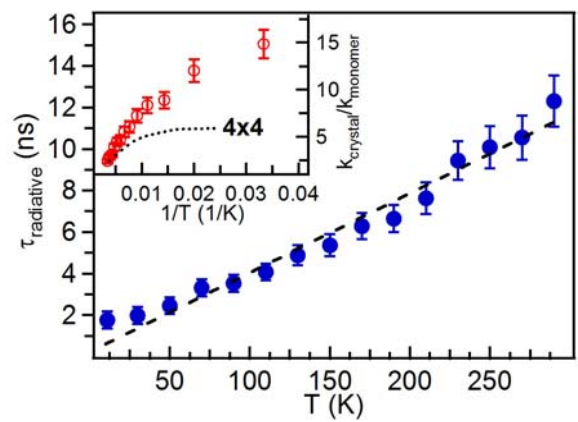

Figure 3, A.Camposeo et al.

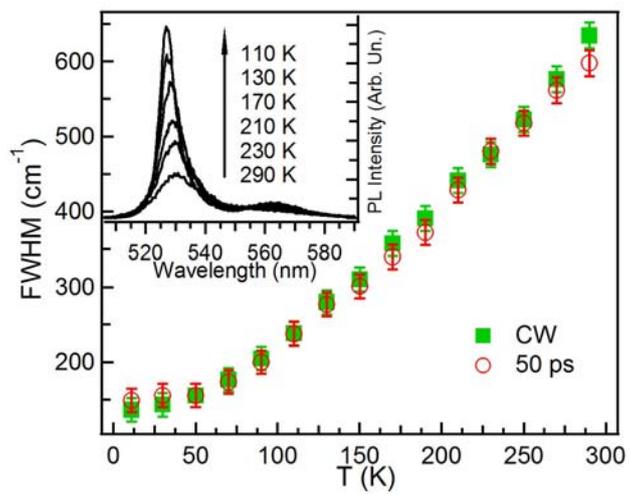

Figure 4, A.Camposeo et al.